\documentclass{iau} 
\usepackage{graphicx}

\title[Green valley galaxies] 
{Properties of green valley galaxies in relation to their selection criteria}

\author[B. Nyiransengiyumva, M. Povi\'c, P. Nkundabakura \& A. Mahoro]    
{Beatrice Nyiransengiyumva$^{1,2}$, Mirjana Povi\'c$^{3,4}$, Pheneas Nkundabakura$^{2}$,
 \and Antoine Mahoro$^{5,6}$}

\affiliation{$^{1}$Mbarara University of Science and Technology (MUST), P.O. Box 1410, Uganda\\ 
$^{2}$University of Rwanda, College of Education, P.O. Box 5039, Kigali, Rwanda\\
$^{3}$Ethiopian Space Science and Technology Institute (ESSTI), Entoto Observatory and Research Center (EORC),
Astronomy and Astrophysics Research and Development Division, P.O. Box 33679, Addis Ababa, Ethiopia\\
$^{4}$Instituto de Astrof\'isica de Andaluc\'ia (IAA-CSIC), Glorieta de la Astronom\'ia s/n, \\ 18008 Granada, Spain\\
$^{5}$South African Astronomical Observatory, P.O. BOX: 9 Observatory, Cape Town, South Africa\\
$^{6}$Department of Astronomy, University of Cape Town, Private Bag X3, Rondebosch 7701, South Africa\\}

\pubyear{2019}
\volume{356}  
\jname{Nuclear activity in galaxies across cosmic time}
\editors{M. Povi\'c, P. Marziani, J. Masegosa, H. Netzer,\\ S. H. Negu, \&
	S. B. Tessema, eds.}
\begin{document}

\maketitle

\begin{abstract}
The distribution of galaxies has been studied to show the difference between the blue cloud and red sequence and to define the green valley region. However, there are still many open questions regarding the importance of the green valley for understanding the morphological transformation and evolution of galaxies, how galaxies change from late-type to early-type and the role of AGN in galaxy formation and evolution scenario. The work focused on studying in more details the properties of green valley galaxies by testing the six most used selection criteria, differences between them, and how they may affect the main results and conclusions.
The main findings  are that, by selecting the green valley galaxies using different criteria, we are selecting different types of galaxies in terms of their stellar masses, sSFR, SFR, spectroscopic classification and morphological properties, where the difference was more significant for colour criteria than for sSFR and SFR vs. M$_{*}$ criteria.
\end{abstract}

\firstsection 
\section{Introduction}

\noindent The bi-modality in the distribution of galaxies usually obtained in colour-mass, colour-magnitude (CMD) or colour-star formation rate diagrams has been studied to show the difference between the blue cloud and red sequence galaxies and to define the intermediate green valley region. In general, blue cloud galaxies are activity star-forming sources that are rich in gas, while red sequence galaxies are mainly abundant quiescent sources and a small fraction of dusty star-forming galaxies and edge-on systems (\cite[Blanton \& Moustakas, 2009]{Blanton2009}). Between the red sequence and the blue cloud there is a sparsely populated green valley region  that has been viewed as the crossroads of galaxy evolution where the galaxies in it are thought to represent the transition population between the blue cloud of star-forming galaxies and the red sequence of quenched and passively evolving galaxies (\cite[Schiminovich et al. 2007]{Schiminovich2007}, \cite[Povíc et al. 2012]{Povi}, \cite[Salim 2014]{Salim2014}, \cite[Lin et al. 2017]{Lin2017}, \cite[Coenda
et al. 2018]{Coenda}, \cite[Bryukhareva \& Moiseev, 2019]{Bryukhareva}, \cite[Phillipps et al. 2019]{Phillipps2019}).\\\\

\noindent Several selection criteria have been used to define green valley in order to study the properties of the green galaxy population, using colours such as U - V (\cite[Brammer et al. 2009]{Brammer2009}), U - B (\cite[Mendez et al. 2011]{Mendez2011}, \cite[Mahoro et al. 2017, 2019]{Mahoro2017, Mahoro2019}), NUV - r (\cite[Wyder et al. 2007]{Wyder2007}, \cite[Lee et al. 2015]{Lee2015}, \cite[Coenda et al. 2018]{Coenda}), g - r (\cite[Trayford et al. 2015, 2017]{Trayf,Trayford}, \cite[Walker et al. 2013]{Walker2013}, \cite[Eales et al. 2018]{Eales2018}), u - r (\cite[Bremer et al. 2018]{Bremer2018}, \cite[Eales et al. 2018]{Eales2018}, \cite[Kelvin et al. 2018]{Kelvin2018}, \cite[Ge et al. 2018]{Ge2018}, \cite[Phillipps et al. 2019]{Phillipps2019}), specific SFR (sSFR) (\cite[Schiminovich et al. 2007]{Schiminovich2007}, \cite[Salim et al. 2009]{Salim2009}, \cite[Salim 2014]{Salim2014}, \cite[Phillipps et al. 2019]{Phillipps2019}, \cite[Starkenburg et al. 2019]{Starkenburg2019}), and the SFR-stellar mass (M$_{*}$) diagram (\cite[Noeske et al. 2007;
Chang et al. 2015]{Noeske2007, Chang/}). Therefore, the study of green valley galaxies provides us crucial clues to connect the red sequence and blue cloud galaxies in terms of their star formation quenching and evolution.

\section{Data and sample selection}
\noindent We used optical and ultraviolet (UV) data in this study. The optical data are from the Sloan Digital Sky Survey (SDSS) Data Release 7 (DR7) with photometric system of five filters named u, g, r, i and z with limiting magnitudes of 22.0, 22.2, 22.2, 21.3 and 20.3 respectively (\cite[York et al. 2000]{York2000}). For SFR, M$_{*}$ and sSFR we used MPA-JHU catalogue (\cite[Kauffmann et al. 2003; Brinchmann et al. 2004; Tremonti et al. 2004]{Kauffmann2003,Brinchmann,Tremonti}). The ultraviolet data are from the GALEX-MIS (GR5) survey with limiting magnitudes of 22.6 and 22.7 in far-UV (FUV) and near-UV (NUV) respectively (\cite[Martin et al. 2005]{Martin}).\\

\noindent Six criteria commonly used in previous studies were tested to select the green valley samples.
\begin{enumerate}
  \item The NUV absolute magnitude minus the r absolute magnitude (M$_\mathrm{NUV}$ - M$_\mathrm{r}$) and the g absolute magnitude minus r absolute magnitude  (M$_\mathrm{g}$ - M$_\mathrm{r}$). The range of green valley galaxies is defined in between 2.8 $<$ M$_\mathrm{NUV}$ - M$_\mathrm{r}$ $<$ 4 and 0.63 $ <$ M$_\mathrm{g}$ - M$_\mathrm{r}$ $<$ 0.75, as shown in Figure~\ref{fig1} which is in line with previous studies (\cite[Belfiore et al. 2017]{Belfiore}, \cite[Taylor et al. 2017]{Taylor2017}, \cite[Eales et al. 2018]{Eales2018}).
\begin{figure}[!h]
\begin{center}
\includegraphics[width=1.85in, height=1.6in]{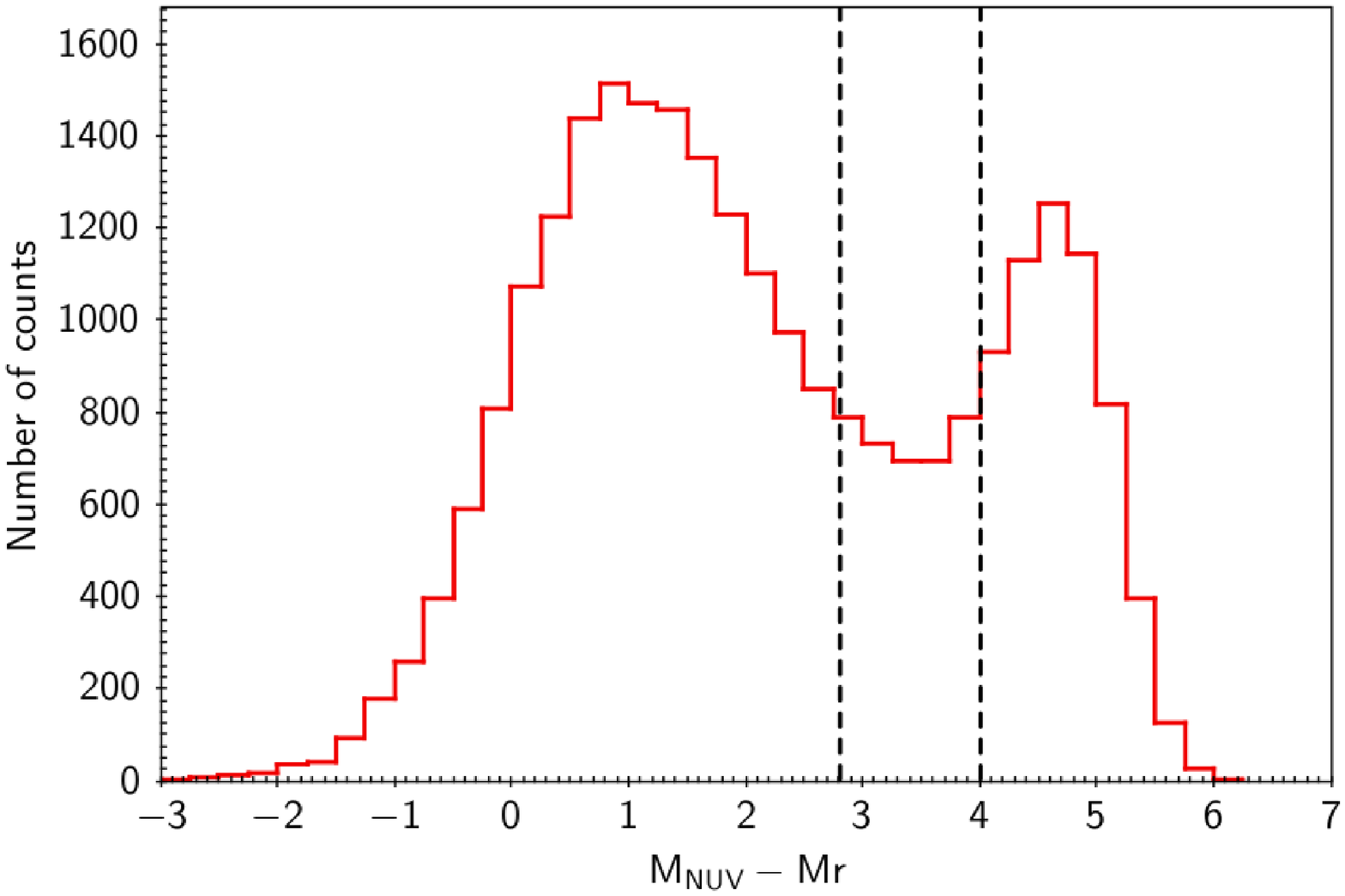}
\includegraphics[width=1.85in, height=1.6in]{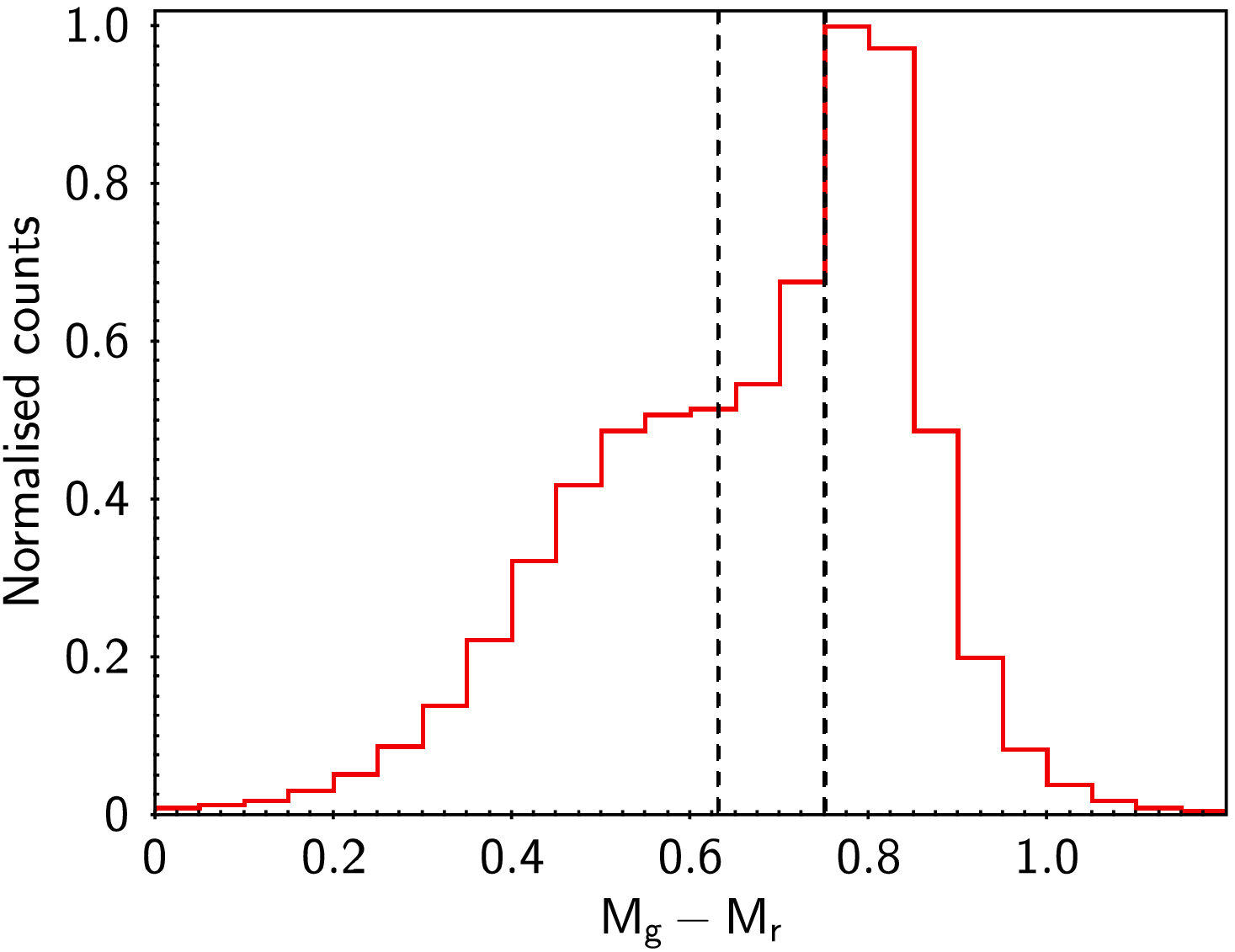}
 \caption{Distributions of M$_\mathrm{NUV}$ - M$_\mathrm{r}$ colour for UV sample (left panel) and M$_\mathrm{NUV}$ - M$_\mathrm{r}$ colour for optical sample (right panel). Green valley region is represented in between the two black dashed lines.}
   \label{fig1}
\end{center}
\end{figure}

\item sSFR in UV and in optical. The green valley sample is defined in the range of -11.6 $<$ sSFR$<$ -10.8 for both criteria as shown in Figure~\ref{fig2} which is in line with \cite[Salim et al. 2009]{Salim2009} and \cite[Salim 2014]{Salim2014}. 

\begin{figure}[!h]
\begin{center}
 \includegraphics[width=1.8in, height=1.55in]{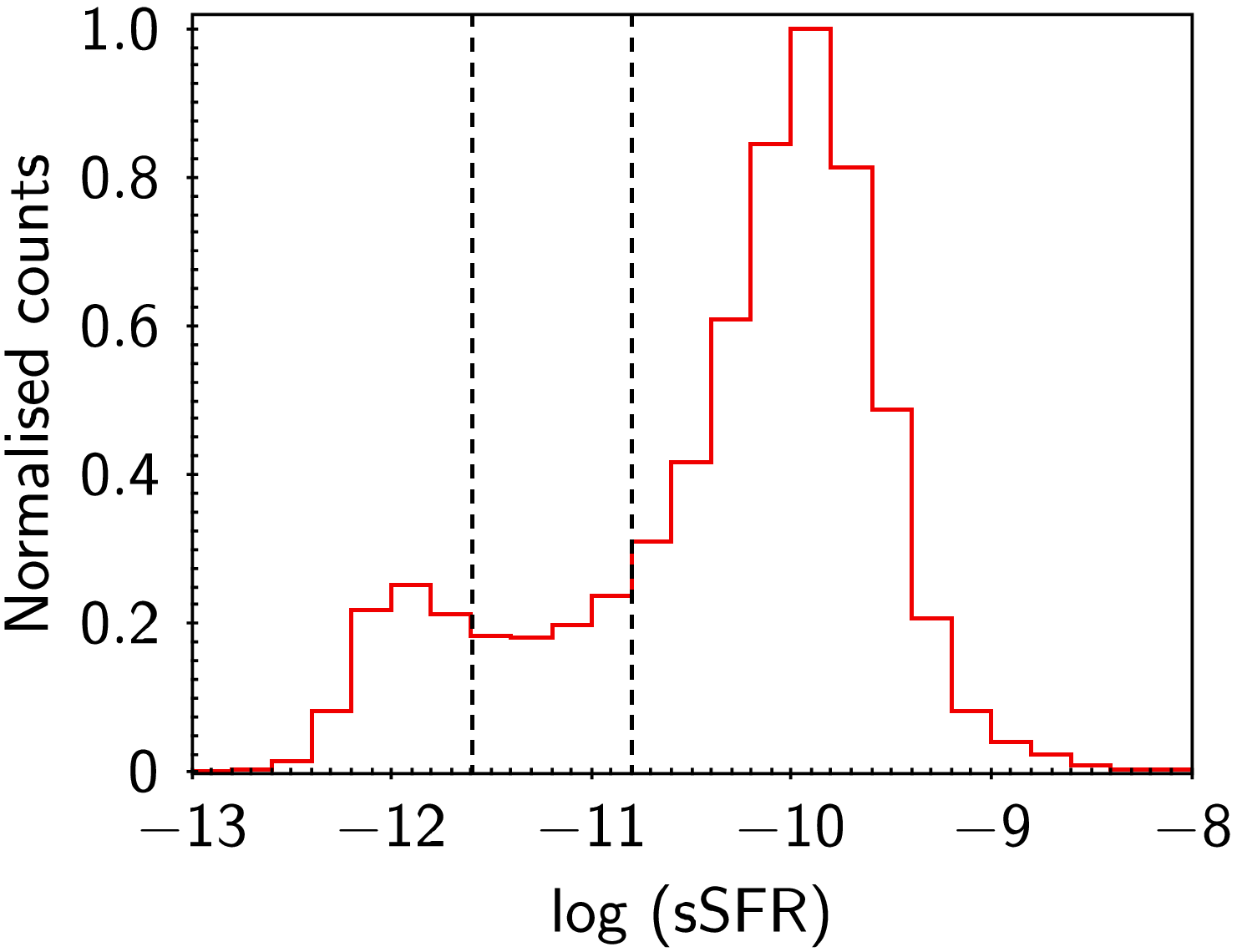} 
 \includegraphics[width=1.8in, height=1.55in]{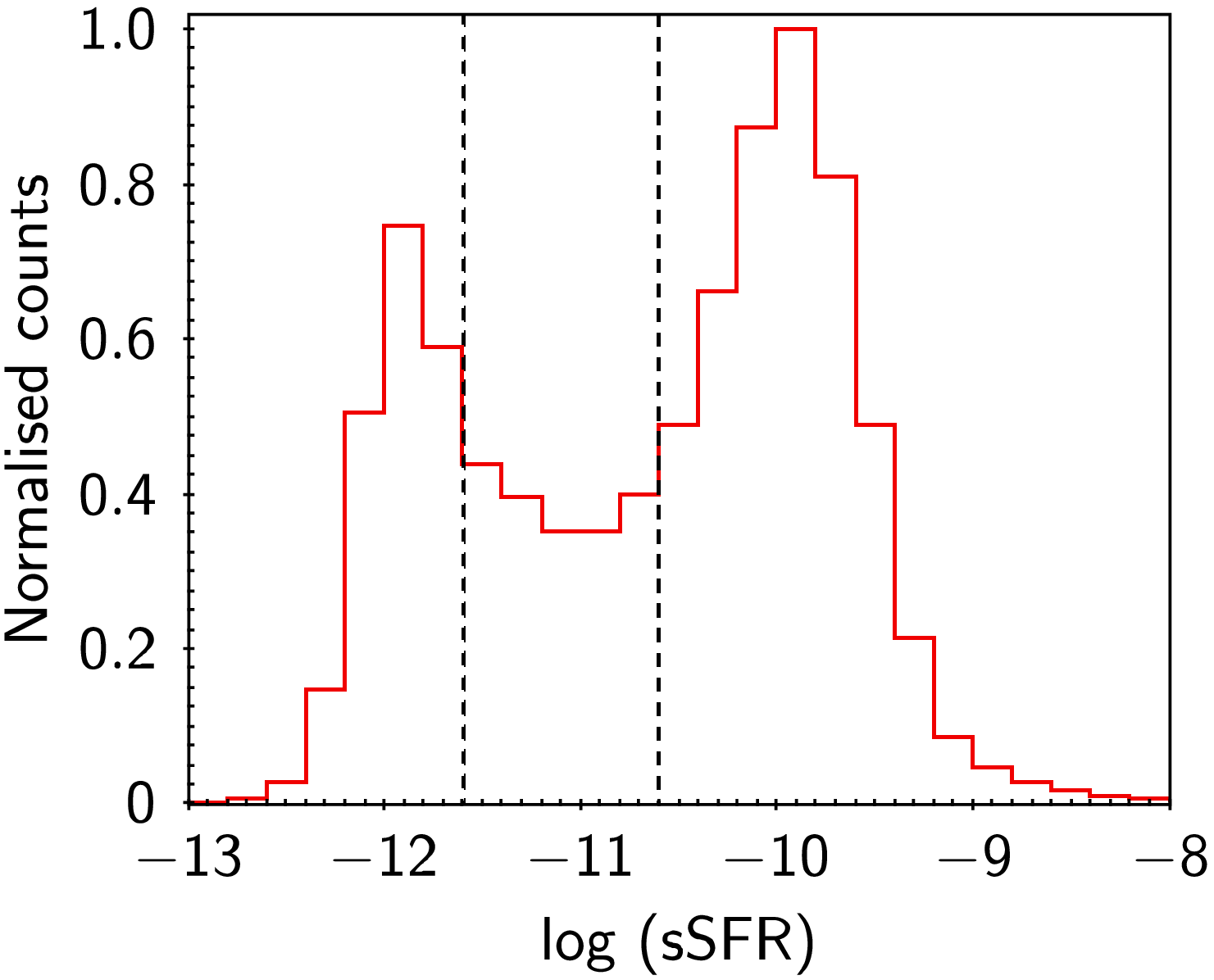} 
 \caption{Distributions of sSFR for UV (left panel) and optical (right panel) samples. Green valley region is represented in between the two black dashed lines. }
   \label{fig2}
\end{center}
\end{figure}
\item SFR vs. M$_\mathrm{*}$ in UV and in optical. The green valley sample is defined as shown in Figure~\ref{fig3} (in line with \cite[Noeske et al. 2007]{Noeske2007}, \cite[Chang et al. 2015]{Chang}).
\begin{figure}[!h]
\begin{center}
\includegraphics[width=2in, height=1.8in]{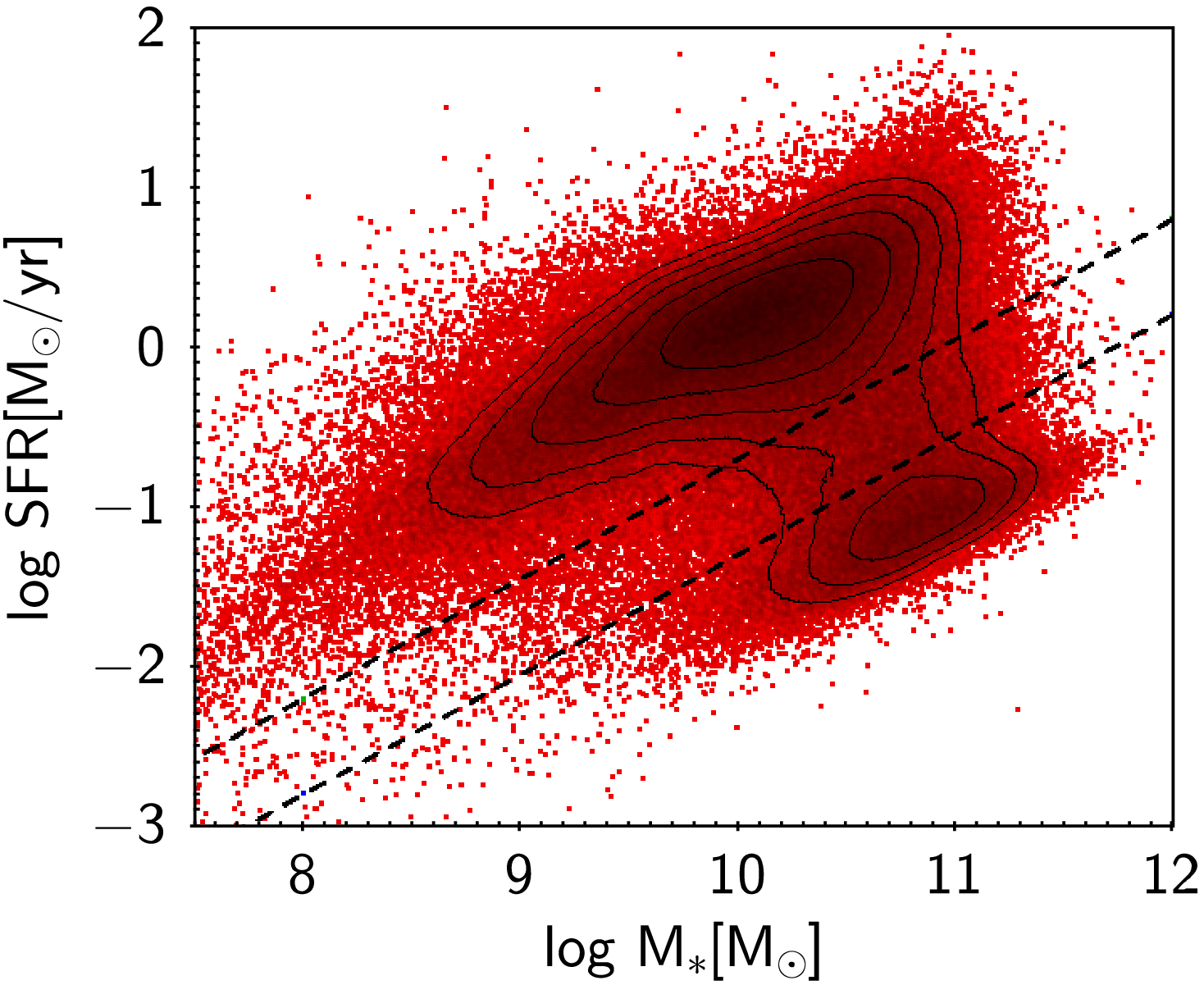} 
\includegraphics[width=2in, height=1.8in]{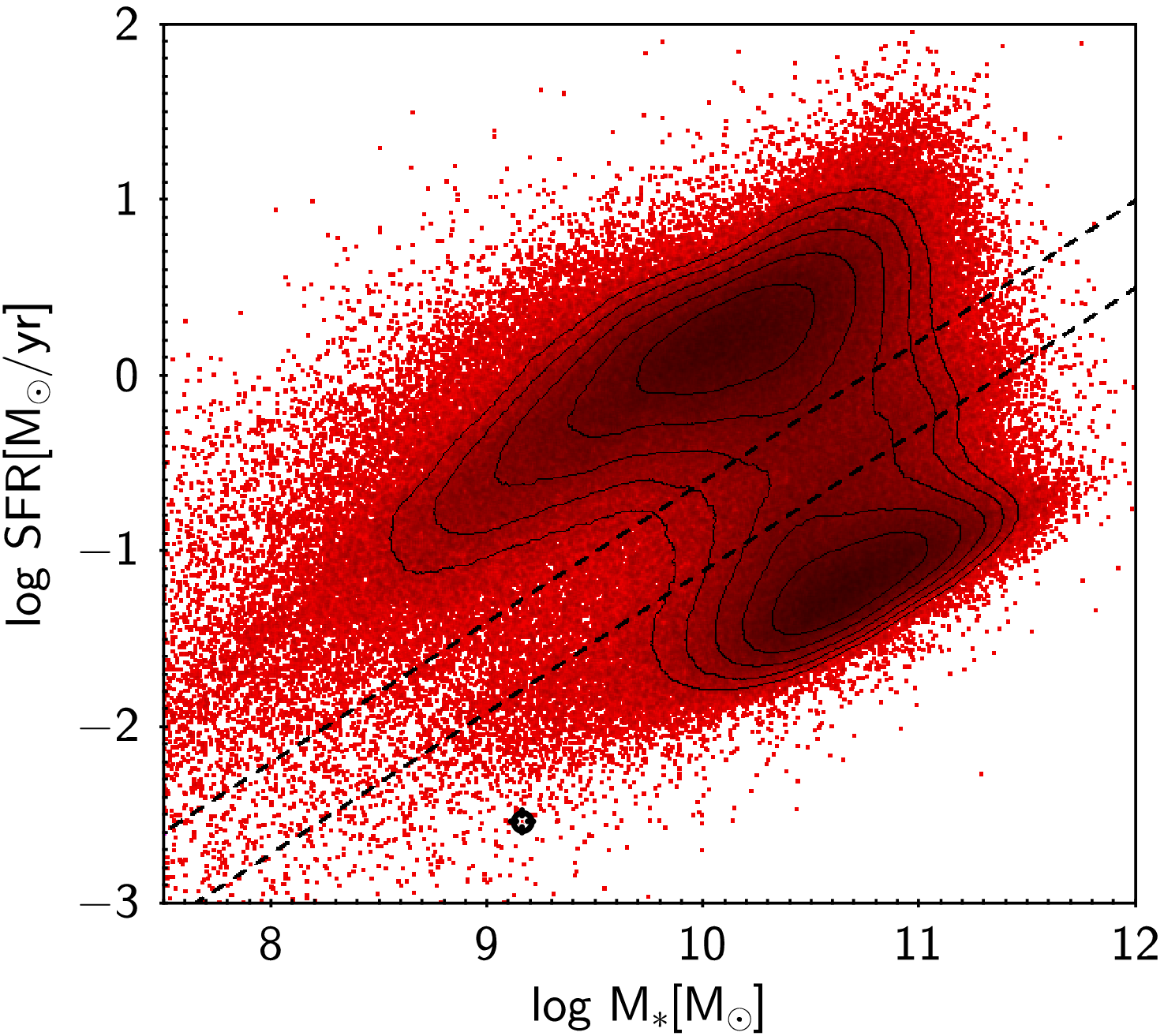} 
 \caption{SFR versus stellar mass for UV sample (left panel) and optical sample (right panel).Green valley region is represented in between the two black dashed lines.}
   \label{fig3}
\end{center}
\end{figure}
\end{enumerate}

\section{Analysis and results}
\begin{figure}[!h]
\begin{center}
\includegraphics[width=2.6in, height=2.5in]{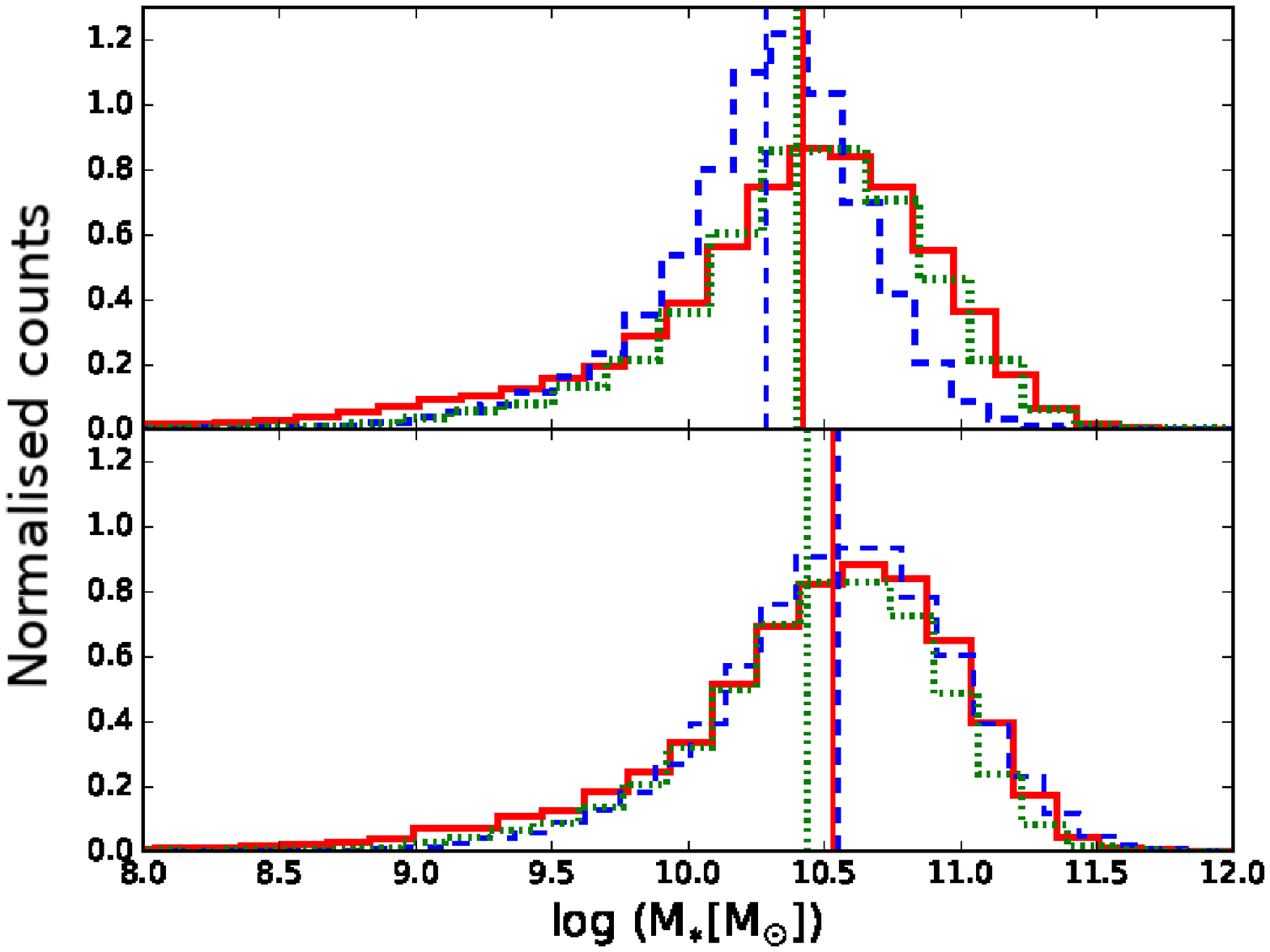}
\includegraphics[width=2.6in, height=2.5in]{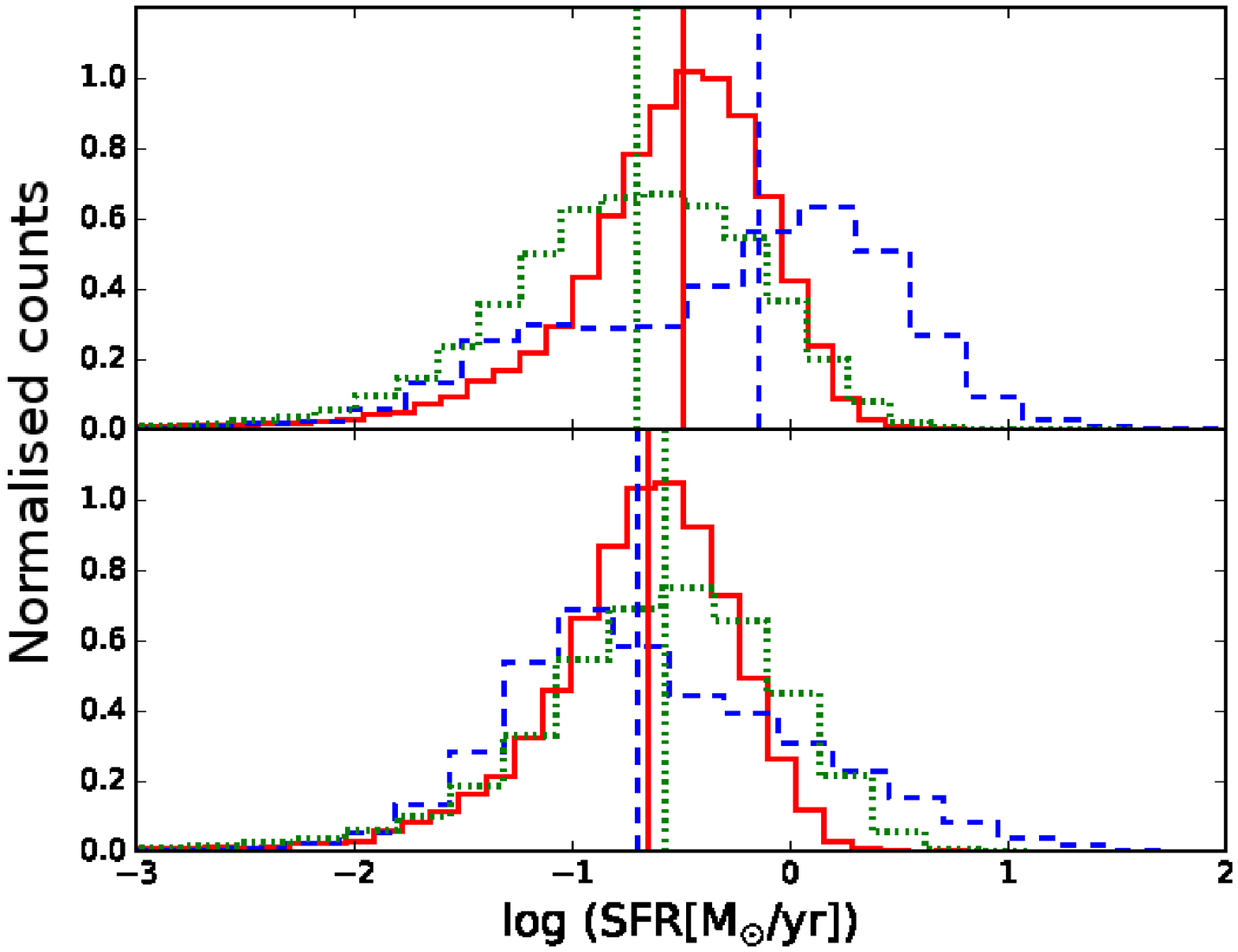}
\includegraphics[width=2.8in, height=2.3in]{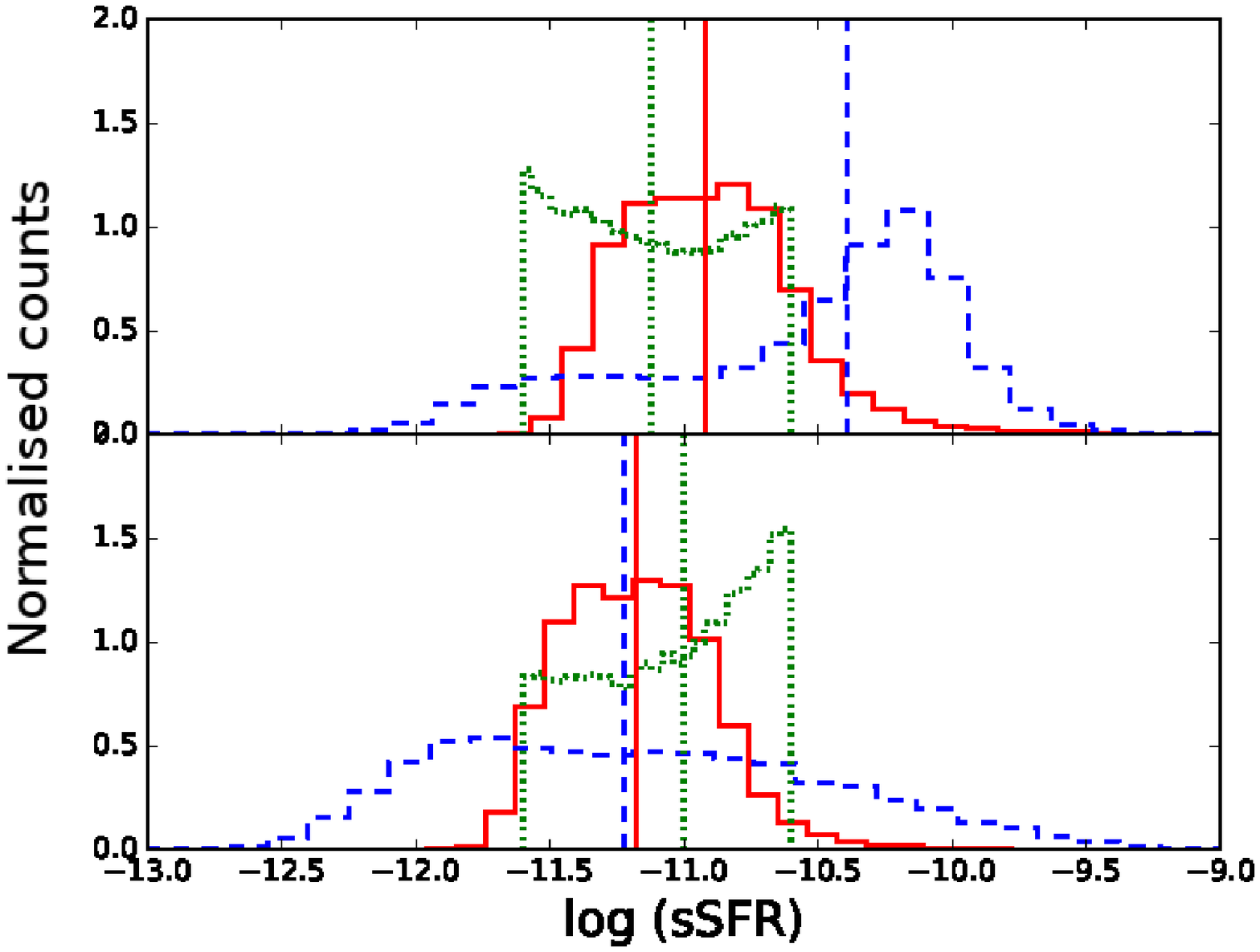}
 \caption{Comparison of stellar mass, SFR, and sSFR in optical (top panels) and UV (bottom panels). The blue dashed, green dotted, and red solid histograms in all figures represent the green valley sample selected using colours (M$_{\rm {NUV}}$-M$_{r}$ and M$_{g}$-M$_{r}$), sSFR, SFR vs. stellar mass criterion. The vertical dashed lines in
all figures represent the median values for each histogram,  where each median has the same colour as the one of the corresponding histogram.}
   \label{fig4}
\end{center}
\end{figure}

\noindent We analysed the distributions of stellar mass, SFR and sSFR in different green valley samples. Figure~\ref{fig4} represents distributions of these parameters, respectively, in optical (top panels) and UV (bottom panels). In each panel, three samples are compared using:  SFR-SM criteria (red solid line), colour criteria (blue dashed line), and  sSFR criteria (green dotted line). It can be seen that when using different criteria we are selecting different galaxies in terms of their stellar mass, where more massive galaxies are selected when using UV data. When observing the total samples, in general SFR and sSFR are higher in optically selected green valley galaxies than in UV, however this difference is more significant when using colour and/or SFR vs. stellar mass criteria.\\
\noindent Using BPT-NII diagram (\cite[Baldwin et al. 1981]{Baldwin}), we classified green valley samples into star forming, composites, Seyfert 2 and LINERs galaxies. Most of the galaxies in the green valley are classified as star forming and Composites where the main difference is when using colour criteria. We found higher fraction of star forming and composite galaxies being selected in optical and UV colour criterion, respectively, while in general higher fraction of Seyfert 2 and LINER galaxies have been selected in UV. \\

\noindent We used the visual morphological classification from the Galaxy Zoo survey (\cite[Lintott et al. 2011]{Lintott}), where galaxies have been classified as elliptical, spiral, and uncertain.When comparing the fraction of elliptical and spiral galaxies selected by different green valley criteria, the only significant difference was found between the two colour criteria where we are detecting more spiral and less elliptical galaxies in optical in comparison to UV (41\% vs. 24\% for spirals, and 3\% vs. 14\% for ellipticals, respectively).

\section{Conclusions}
\noindent This work conducted a study on properties of galaxies in different green valley samples selected based on different criteria and analysed the differences and similarities
from one criterion to another. Our main findings are:\\
\begin{itemize}
 \item By selecting the green valley galaxies based on UV and optical data, we are selecting different types of galaxies in terms of their stellar
masses, SFRs, sSFRs, morphological classification, spectroscopic types etc.
\item With colour criteria, there is a difference in terms that with optical colour we are selecting much more spirals than ellipticals but for UV colour criteria, we have more ellipticals. For sSFR and SFR-M, the difference is not more
pronounced.
\item With spectroscopic types, more star forming galaxies were selected
in optical sample while composites and AGNs were selected more in
UV sample. This is very pronounced for colour criteria.
\end{itemize}

\section{Acknowledgements}
\noindent Financial support from the Swedish International Development Cooperation Agency (SIDA) through the International Science  Programme (ISP) to the East African Astrophysics Research Network (EAARN) is gratefully acknowledged. MP acknowledges financial support from the Ethiopian Space Science and Technology Institute (ESSTI) under the
Ethiopian Ministry of Innovation and Technology (MoIT),
and support from the Spanish Ministry of Science, Innovation and Universities
(MICIU) through project
AYA2016-76682C3-1-P.
AM acknowledges support from the National Research Foundation of South Africa (Grant Numbers 110816)



\begin{thebibliography}{}
\bibitem[Baldry et al. (2004)]{Baldry2004}
{Baldry I. K., Glazebrook K., Brinkmann J., Ivezi ́c Ž., Lupton R. H., Nichol
R. C., Szalay A. S.} 2004,
\textit{ApJ}, 600, 681

\bibitem[Baldwin et al. (1981))]{Baldwin}
{Baldwin J. A., Phillips M. M., Terlevich R.} 1981, 
\textit{PASP}, 93, 5

\bibitem[Belfioree et al. 2017]{Belfiore}
{Belfiore F., et al., 2017, preprint,} 
\textit{(arXiv:1710.05034)}

\bibitem[Blanton \& Moustakas, (2009)]{Blanton2009}
{Blanton M. R., Moustakas J.}  2009,
\textit{ARA\&A}, 47, 159

\bibitem[Brammer et al. 2009]{Brammer2009}
{Brammer G. B., et al.} 2009, 
\textit{ ApJ}, 706, L173

\bibitem[Bremer et al. 2018]{Bremer2018}
{Bremer M. N., et al.} 2018, 
\textit{MNRAS}, 476, 12

\bibitem[Brinchmann et al. (2004)]{Brinchmann}
{Brinchmann J., Charlot S., White S. D. M., Tremonti C., Kauffmann G.,
Heckman T., Brinkmann J.} 2004, 
\textit{MNRAS}, 351, 1151

\bibitem[Bryukhareva \& Moiseev, (2019)]{Bryukhareva}
{Bryukhareva T. S., Moiseev A. V.} 2019, 
\textit{MNRAS}, 489, 3174.
\bibitem[Chang et al. (2015)]{Chang}
{Chang Y.-Y., van der Wel A., da Cunha E., Rix H.-W.} 2015, 
\textit{ ApJS}, 219, 8.

\bibitem[Coenda et al. (2018)]{Coenda}
{Coenda V., Martínez H. J., Muriel H.} 2018, 
\textit{PASP}, 473, 5617.

\bibitem[Eales et al. 2018]{Eales2018}
{Eales S. A., et al.} 2018, 
\textit{MNRAS}, 481, 1183.
\bibitem[Ge et al. 2019]{Ge2018}
{Ge X., Gu Q.-S., Chen Y.-Y., Ding N} 2019, 
\textit{Research in Astronomy and
Astrophysics}, 19, 027

\bibitem[Kauffmann et al. 2003]{Kauffmann2003}
{Kauffmann G., et al} 2003, 
\textit{MNRAS}, 341, 33

\bibitem[Kelvin et al. 2018]{Kelvin2018}
{Kelvin L. S., et al.} 2018, 
\textit{MNRAS}, 477, 4116

\bibitem[Kewley et al. (2001)]{Kewl}
{Kewley L. J., Dopita M. A., Sutherland R. S., Heisler C. A., Trevena J.} 2001, 
\textit{ApJ}, 556, 121.

\bibitem[Lee et al. (2015)]{Lee2015}
{Lee G.-H., Hwang H. S., Lee M. G., Ko J., Sohn J., Shim H., Diaferio A} 2015, 
\textit{ApJ}, 800, 80.

\bibitem[Lin et al. (2017)]{Lin2017}
{Lin L., et al.} 2017, 
\textit{ApJ}, 851, 18. 

\bibitem[Lintott et al. (2011)]{Lintott}
{Lintott C., et al.} 2011, 
\textit{MNRAS}, 410, 166.

\bibitem[Mahoro et al. (2017)]{Mahoro2017}
{Mahoro A., Povi\'c M., Nkundabakura P.} 2017, 
\textit{MNRAS}, 471, 3226

\bibitem[Mahoro et al. 2019]{Mahoro2019}
{Mahoro A., Povi\'c M., Nkundabakura P., Nyiransengiyumva B., Väisänen
P.} 2019,
\textit{MNRAS}, 485, 452

\bibitem[Martin et al. 2005]{Martin}
{Martin D. C., et al.} 2005,
\textit{ApJ}, 619, L1

\bibitem[Mendez et al. 2011]{Mendez2011}
{Mendez A. J., Coil A. L., Lotz J., Salim S., Moustakas J., Simard L} 2011,
\textit{ApJ}, 736, 110


\bibitem[Noeske et al. 2007]{Noeske2007}
{Noeske K. G., et al.} 2007,
\textit{ApJ}, 660, L43
\bibitem[Phillipps et al. (2019)]{Phillipps2019}
{Phillipps S., et al.} 2019,
\textit{MNRAS}, 485, 5559

\bibitem[Povi\'c et al. (2012)]{Povi}
{Povi\'c M., et al.} 2012, 
\textit{A\&A}, 541, A118

\bibitem[Salim  (2014)]{Salim2014}
{Salim S., 2014} 1995,
\textit{Serbian Astronomical Journal}, 181, 1 


\bibitem[Salim et al. 2009]{Salim2009}
{Salim S., et al.} 2009,
\textit{ApJ}, 700, 161 

\bibitem[Schawinski et. al (2007)]{Scha}
{Schawinski K., Thomas D., Sarzi M., Maraston C., Kaviraj S., Joo S.-J., Yi
S. K., Silk J.} 2007, 
\textit{MNRAS}, 382, 1415.

\bibitem[Schiminovich et al. 2007]{Schiminovich2007}
{Schiminovich D., et al.} 2007, 
\textit{ApJS}, 173, 315

\bibitem[Starkenburg et al. 2019]{Starkenburg2019}
{Starkenburg T. K., Tonnesen S., Kopenhafer C.} 2019, 
\textit{ApJ}, 874, L17
\bibitem[Trayford et al. 2017]{Trayford}
{Trayford J. W., et al., 2017, preprint} 
\textit{(arXiv:1705.02331)}

\bibitem[Trayford et al. (2015)]{Trayf}
{Trayford J. W., et al.} 2015, 
\textit{MNRAS}, 452, 2879

\bibitem[Tremonti et al. (2004)]{Tremonti}
{Tremonti C. A., et al.} 2004, 
\textit{ApJ}, 613, 898
\bibitem[Walker et al. 2013]{Walker2013}
{Walker L. M., et al.} 2013,
\textit{ApJ}, 775, 129


\bibitem[Wyder et al. 2007 ]{Wyder2007}
{Wyder T. K., et al.} 2007,
\textit{ApJS}, 173, 293



\bibitem[York et al. 2000]{York2000}
{York D. G., et al}  2000,
\textit{AJ}, 120, 1579

\end{thebibliography}
\end{document}